\documentclass[twocolumn,twoside]{revtex4}
\usepackage{graphicx}
\usepackage{fancyhdr}

\usepackage[T1]{fontenc}
\usepackage{microtype}
\usepackage[textsize=scriptsize,backgroundcolor=red!70,linecolor=red]{todonotes}
\usepackage{booktabs}
\usepackage{amsmath}
\usepackage{amsfonts}
\usepackage{amssymb}
\usepackage{hyperref}
\usepackage[all]{hypcap}
\usepackage{paralist}
\usepackage{multirow}
\usepackage{graphicx}
\usepackage{dcolumn}
\usepackage{bm}
\usepackage{epsf}

\newcommand*{\rom}[1]{\expandafter\@slowromancap\romannumeral #1@}

\def\be{\begin{equation}}
\def\ee{\end{equation}}
\def\bea{\begin{eqnarray}}
\def\eea{\end{eqnarray}}
\def\gsim{\ \rlap{\raise 2pt\hbox{$>$}}{\lower 2pt \hbox{$\sim$}}\ }
\def\lsim{\ \rlap{\raise 2pt\hbox{$<$}}{\lower 2pt \hbox{$\sim$}}\ }
\def\dslash{\kern-4pt \not{\hbox{\kern-2pt $\partial$}}}
\def\pslash{\not{\hbox{\kern-2pt p}}}

\def\beq{\begin{equation}}
\def\eeq{\end{equation}}

\def \be{\beta}

\def \({\left(}
\def \){\right)}
\def \[{\left[}
\def \]{\right]}

\def\beq{\begin{equation}}
\def\eeq{\end{equation}}
\def\beqa{\begin{eqnarray}}
\def\eeqa{\end{eqnarray}}

\def\ra{\rightarrow}

\def\beq{\begin{equation}}
\def\eeq{\end{equation}}
\def\beqa{\begin{eqnarray}}
\def\eeqa{\end{eqnarray}}

\begin{document}

	\title{Massive Right-handed Neutrinos in B Decays}

%

\author{Hongkai Liu\\}
\affiliation{Department of Physics, Technion – Israel Institute of Technology, Haifa 3200003, Israel}

\begin{abstract}
In this paper, we present the  differential decay distributions
 for $\bar B \to  D^{(*)} \ell \bar{X}$  decays with a massive right-handed neutrino in the low-energy effective field theory framework and show how the massive effects of the RH neutrinos can explain the positive value of the difference in forward-backward asymmetries, $\Delta A_{\text{FB}}\equiv A_{\text{FB}}^\mu-A_{\text{FB}}^e$, tentatively inferred from Belle data. We also make predictions for $q^2$ dependent  angular observables to motivate future measurements. 
\end{abstract}

\maketitle

\thispagestyle{fancy}


\section{Introduction}
The recently noticed $4~\sigma$ deviation form the standard model (SM) predictions in the difference of the forward-backward asymmetry between the muon and electron channel  ($\Delta A_{FB} \equiv A_{FB}^\mu - A_{FB}^e$) measured from Belle data in Ref.~\cite{Belle:2018ezy} motivates the new physics (NP) in the electron and muon sector. The anomalies in the 
$\Delta$ observables are quite interesting, as they have very little form factor uncertainties and hence any measured deviations from the SM predictions for these observables would be clear signs of NP.

Massive right-handed (RH) or sterile neutrinos are well-motived hypothetical particles to explain many phenomena beyond the standard model (BMS), such as neutrino oscillations and dark matter. RH neutrinos are sterile under the SM gauge interactions and can be incorporated into the standard model effective field theory (SMEFT).  The resulting EFT~\cite{delAguila:2008ir, Aparici:2009fh, Bhattacharya:2015vja, Liao:2016qyd, Bischer:2019ttk}, called SMNEFT, includes additional interactions of the RH neutrinos with SM fields. The mass scale of the RH neutrino can vary over a large range. We consider the case of a light RH neutrino so that it appears as an explicit degree of freedom in the EFT framework. 
 The differential decay distribution for $\bar B\ra D \ell \bar \nu$ with a massless RH neutrino is given in Ref.~\cite{Mandal:2020htr}. We generalize the result for a nonzero RH neutrino mass $m_N$. A finite $m_N$ affects both the phase space and the leptonic helicity amplitudes.
 Given the anomalies in the measured value of 
$(g-2)_\mu$~\cite{Muong-2:2021ojo} and neutral-current $b \to s \mu^+\mu^-$ decays~\cite{LHCb:2021trn}, we assume the massive RH neutrinos can be produced from B meson decays and couple to the muon sector to explain the anomaly in $\Delta A_{FB}$. 

\section{Framework}
The $B$ charged-current decay $b\rightarrow c\ell\bar{X}$ can be described by the operators in the low-energy effective field theory (LEFT)
\beq
-{\cal L}_{\text{eff}} = \frac{4G_FV_{cb}}{\sqrt{2}}(\mathcal{O}^V_{LL}+\sum_{\substack{X=S,V,T\\\alpha,\beta=L,R}}C^X_{\alpha\beta}~\mathcal{O}^X_{\alpha\beta})\,,
\label{eq:wc}
\eeq
where
\beqa
\mathcal{O}^V_{\alpha\beta} &\equiv& (\bar{c}\gamma^\mu P_\alpha b) (\bar{\ell}\gamma^\mu P_\beta X)\,,\\
\mathcal{O}^S_{\alpha\beta} &\equiv& (\bar{c} P_\alpha b) (\bar{\ell} P_\beta X)\,,\\
\mathcal{O}^T_{\alpha\beta} &\equiv& \delta_{\alpha\beta} (\bar{c} \sigma^{\mu\nu} P_\alpha b) (\bar{\ell}\sigma_{\mu\nu} P_\beta X)\,,
\eeqa
with $X$ the left-handed (LH) SM neutrinos or RH neutrinos.
The SM and NP contributions are in the first and second terms in Eq.~(\ref{eq:wc}), respectively.
After matching at the electroweak scale, only the operators $\mathcal{O}^V_{LL},\,\mathcal{O}^S_{LL},\,\mathcal{O}^S_{RL}$, and $\mathcal{O}^T_{LL}$ can arise from the four-fermion operators in SMEFT, while SMNEFT yields four more operators: $\mathcal{O}^V_{RR},\,\mathcal{O}^S_{LR},\,\mathcal{O}^S_{RR}$, and $\mathcal{O}^T_{RR}$; see Table~\ref{Table: op}. Note that $\mathcal{O}^V_{LR}$ and $\mathcal{O}^V_{RL}$ cannot be produced from the four-fermion operators in SMNEFT. 
\begin{table}
	\centering
 	\caption{The origin of low-energy effective operators from SMNEFT.}
	\begin{tabular}{| c   c   c   c || c  c  c c | }
		\toprule
		$\mathcal{O}^{(3)}_{\ell q}$  & $\mathcal{O}^{(1)}_{\ell equ}$& $\mathcal{O}_{\ell edq}$  & $\mathcal{O}^{(3)}_{\ell eqd}$ & $\mathcal{O}_{nedu}$  & $\mathcal{O}_{\ell nuq}$& $\mathcal{O}^{(1)}_{\ell nqd}$  & $\mathcal{O}^{(3)}_{\ell nqd}$   \\
		\midrule
		$\mathcal{O}^V_{LL}$  & $\mathcal{O}^S_{LL}$& $\mathcal{O}^S_{RL}$  & $\mathcal{O}^T_{LL}$ & $\mathcal{O}^V_{RR}$  & $\mathcal{O}^S_{LR}$& $\mathcal{O}^S_{RR}$  & $\mathcal{O}^T_{RR}$    \\
		\midrule
	\end{tabular}
	\label{Table: op}
\end{table}
The renormalization group running of the operators from $\Lambda$ to $m_Z$ and then down to the $m_b$ scale has been discussed in Refs.~\cite{Datta:2020ocb,Datta:2021akg}. In what follows, we work in the LEFT framework keeping in mind that the corresponding SMNEFT WCs can be obtained by carrying out the running and matching. 

\section{Phenomenology}
The differential decay distribution  for $\bar B\ra D \ell \bar X$  can be expressed in terms of the three $\mathcal{J}$ functions as    
\beqa
\frac{d^2\Gamma_{D}}{dq^2d\cos\theta_{\ell}} &=& \mathcal{J}_0(q^2) + \mathcal{J}_1(q^2)\cos\theta_{\ell} + \mathcal{J}_2(q^2)\cos^2\theta_{\ell}\,,\nonumber\\
\eeqa
with $q^2 \equiv (p_{\ell}+p_{\bar N})^2$ and $\theta_{\ell}$ the angle between the charged lepton momentum in the  $\ell\bar X$ rest frame
and the direction of the $D$ momentum in the $\bar B$ rest frame. The differential decay width and angular observable $A_{FB}$ can be written in terms of $\mathcal{J}$ functions
\beqa
\Gamma^D_f(q^2)&\equiv&\frac{d\Gamma_{D}}{dq^2} = 2 \mathcal{J}_0(q^2) + \frac{2}{3} \mathcal{J}_2(q^2),\\
A^D_{FB}(q^2) &\equiv& - \frac{\mathcal{J}_1(q^2)}{d\Gamma_{D}/dq^2}.
\eeqa

\begin{figure}[h]
	\centering
	\includegraphics[width=80mm]{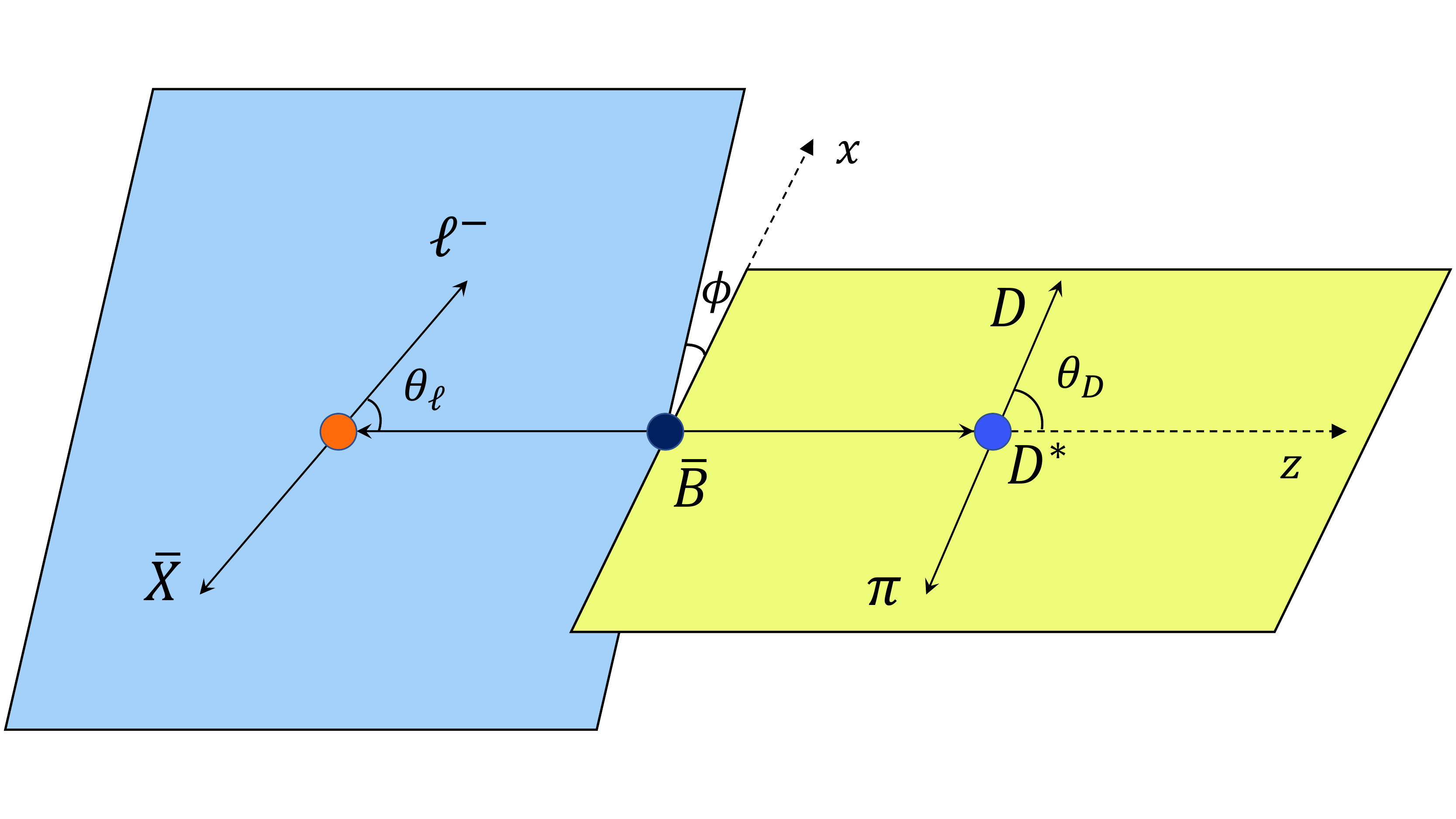}
	\caption{Kinematic variables for $\bar B\ra \ell^-\bar X D^*(\ra D\pi)$.} \label{fig:decay}
\end{figure}

Similarly, the differential decay distribution for $\bar B\ra D^* (\ra D\pi)\ell \bar X$ with nonzero $m_N$, can be written in terms of the 12 different angular structures that appear in the massless RH neutrino case:
\begin{widetext}
\beqa
\frac{8\pi}{3} \frac{d^4\Gamma_{D^*}}{dq^2d\cos\theta_{\ell}d\cos\theta_Dd\phi} &=& (\mathcal{I}_{1s}+\mathcal{I}_{2s}\cos2\theta_{\ell}+ \mathcal{I}_{6s}\cos\theta_{\ell})\sin^2\theta_D \nonumber\\
&+&(\mathcal{I}_{1c}+\mathcal{I}_{2c}\cos2\theta_{\ell}+ \mathcal{I}_{6c}\cos\theta_{\ell})\cos^2\theta_D \nonumber\\
&+&(\mathcal{I}_3\cos2\phi+\mathcal{I}_9\sin2\phi )\sin^2\theta_D \sin^2\theta_{\ell}\\
&+&(\mathcal{I}_4\cos\phi+\mathcal{I}_8\sin\phi )\sin2\theta_D \sin2\theta_{\ell}\nonumber\\
&+&(\mathcal{I}_5\cos\phi+\mathcal{I}_7\sin\phi )\sin2\theta_D \sin\theta_{\ell}\nonumber\,,
\label{Dstardist}
\eeqa
\end{widetext}
where the three angles are defined in Fig.~\ref{fig:decay}. The expression of 
$\mathcal{J}$ and $\mathcal{I}$ functions are quite lengthy as they depend on the WCs, mass of RH neutrinos, $q^2$, and hadronic form factors. We do not present the details of those 
angular functions in this paper for brevity. For the complete expression see {\it Supplemental Material} in Ref.~\cite{Datta:2022czw}.
We adopt the hadronic form factors of Ref.~\cite{Bordone:2019guc} including the corrections up to $1/m_c^2$ in the heavy-quark limit. The $q^2$ distributions of differential decay width and angular observables are related to the angular functions as follow
\beqa
\Gamma^{D^*}_f(q^2)\equiv\frac{d\Gamma_{D^*}}{dq^2} &=& 2\mathcal{I}_{1s}(q^2) +\mathcal{I}_{1c}(q^2) \nonumber\\
&&-\frac{1}{3} (2 \mathcal{I}_{2s}(q^2)+\mathcal{I}_{2c}(q^2))\,.\nonumber\\
A^{D^*}_{FB} (q^2) &=& -\frac{\mathcal{I}_{6s}(q^2) + \frac{1}{2} \mathcal{I}_{6c}(q^2)  }{\Gamma^{D^*}_f(q^2)}\,, \\
F_{L}(q^2) &=& \frac{\mathcal{I}_{1c}(q^2) - \frac{1}{3} \mathcal{I}_{2c}(q^2)  }{\Gamma^{D^*}_f(q^2)}\,,\\
\tilde{F}_{L}(q^2) &=& \frac{1}{3} - \frac{8}{9}  \frac{2 \mathcal{I}_{2s}(q^2) + \mathcal{I}_{2c}(q^2)  }{\Gamma^{D^*}_f(q^2)}\,,\\
S_i(q^2) &=& \frac{\mathcal{I}_i(q^2)}{\Gamma^{D^*}_f(q^2)}.
\eeqa

To compare with experimental measurements, we define 9 bins of the normalized $q^2$-distributions~\cite{Bobeth:2021lya},
\beqa
\Delta x^{D^{(*)}}_{i} \equiv \frac{1}{\Gamma_{\text{tot}}^{D^{(*)}}}\int_{q^2_{i-1}}^{q^2_{i}}dq^2 \Gamma^{D^{(*)}}_f(q^2),\quad i = 2~\text{to}~10\,,
\eeqa
where $\Gamma^{D^{(*)}}_{\text{tot}}$ is the total decay width after integrating $\Gamma^{D^{(*)}}_f(q^2)$ over the entire range of $q^2$. The $q^2$ bins are defined by
\beq
q^2_i \equiv m_B^2 + m_{D^{(*)}}^2-2 m_ B m_{D^{(*)}} \omega_i\,,\quad i= 1~\text{to}~10\,,
\eeq 
with $\omega_i = 1 + i/20$.
The $q^2$-binned and -averaged observables are defined by
\beqa
O_i &\equiv& \frac{1}{\Gamma_{\text{tot}}^{D^{(*)}}}\int_{q^2_{i-1}}^{q^2_{i}}dq^2O(q^2)\Gamma^{D^{(*)}}_f(q^2),\quad i = 2~\text{to}~10\,,\nonumber\\\\
\langle O\rangle&\equiv& \frac{1}{\Gamma^{D^{(*)}}_{\text{tot}}}\int_{q^2_{\text{min}}}^{q^2_{\text{max}}}dq^2 O(q^2)\Gamma^{D^{(*)}}_f(q^2)\,.
\eeqa
The values of $\langle A^{D^*}_{\text{FB}}\rangle, \langle \tilde{F}_L\rangle, \langle F_L\rangle$ and $\langle S_3\rangle$, measured by the Belle experiment are listed in Table~\ref{tab:obs_meas}. 
Measurements of the two ratios of branching fractions $R^{\mu/e}_{D^{(*)}}$
are also listed in Table~\ref{tab:obs_meas}.

\begin{table}[t]
	\begin{center}
 	\caption{Ten observables that are sensitive to NP in the $\mu$ sector. The corresponding predictions for the three BPs of Table~\ref{tab:bmp} are provided.}
		\begin{tabular}{|c|c|c|c|c|}
			\hline\hline
			Observable  & Measurement & BP1 & BP2 & BP3 \\
			\hline
			$\Delta \langle A^{D^*}_{\text{FB}}\rangle$ & $0.0349 \pm 0.0089$ &0.0188&  -0.0014 & -0.0016 \\
			$\Delta \langle F_{L}\rangle $ & $ -0.0065 \pm 0.0059$ & -0.0057&-0.0063&-0.0025 \\
			$\Delta\langle \tilde{F}_{L}\rangle$ & $-0.0107 \pm 0.0142$&-0.0314&-0.0099&-0.0034 \\
			$\Delta\langle S_{3}\rangle$ & $-0.0127 \pm 0.0109$ & 0.0035&0.0049&0.0007 \\
			$R_{D}^{\mu/e}$ & $0.995 \pm 0.022 \pm 0.039$ &1.015&1.036&1.012 \\
			$R_{D^*}^{\mu/e}$  & $0.99 \pm 0.01 \pm 0.03$ &0.983& 1.021& 0.991 \\
			\hline\hline
			$\Delta x_2^{D^*}$ & $-0.0040 \pm 0.0029$ & -0.0153 & -0.0022 & -0.0002\\
			$\Delta x_3^{D^*}$ & $-0.0025 \pm 0.0033$ & 0.0 & -0.0022 & 0.0001\\
			$\Delta x_4^{D^*}$ & $0.0024 \pm 0.0038$ & 0.0014 & -0.0022 & 0.0002\\
			$\Delta x_5^{D^*}$ & $0.0043\pm 0.0046$ & 0.0022 & -0.0006 & 0.0002\\
			$\Delta x_6^{D^*}$ & $-0.0035 \pm 0.0052$ & 0.0027 & 0.0009 & 0.0003\\
			$\Delta x_7^{D^*}$ & $0.0066 \pm 0.0056$ & 0.0030 & 0.0018 & 0.0003\\
			$\Delta x_8^{D^*}$ & $-0.0103 \pm 0.0054$ & 0.0032 &  0.0021 & 0.0003\\
			$\Delta x_9^{D^*}$ & $0.0 \pm 0.0052$ & 0.0031 & 0.0020 & 0.0003\\
			$\Delta x_{10}^{D^*}$ & $0.0019 \pm 0.0044$ & 0.0028 & 0.0017 & 0.0003\\
			\hline\hline
			$\Delta\langle A^D_{\text{FB}}\rangle$ & - & 0.0401&-0.0032&-0.0209\\
			$\Delta\langle S_4\rangle$ & - & 0.0121&0.0087& 0.0021 \\
			$\Delta\langle S_5\rangle$ & - & -0.0128&-0.0051& 0.0015 \\
			\hline\hline
		\end{tabular}
	\end{center}
	\label{tab:obs_meas}
\end{table}

\begin{figure}[t]
\includegraphics[width=0.47\textwidth]{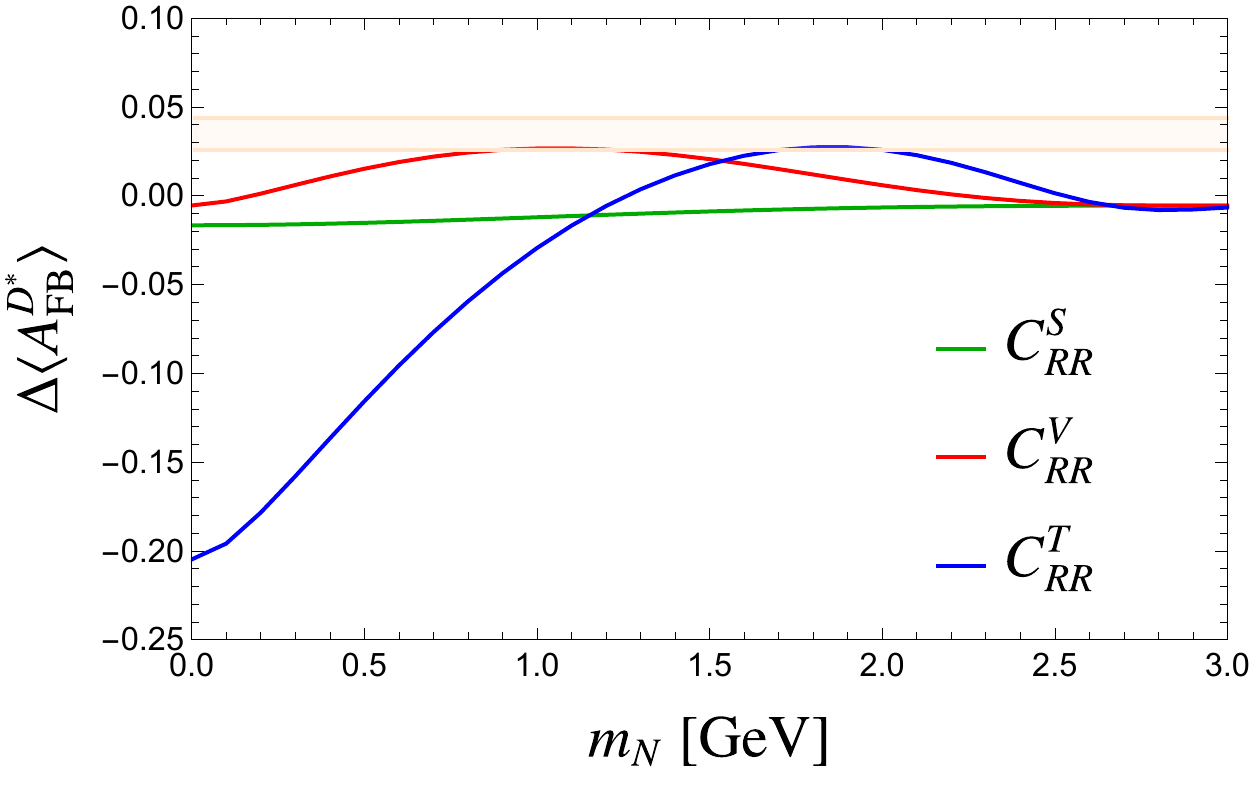}
\caption{$\Delta \langle A^{D^*}_{\text{FB}}\rangle$ as a function of $m_N$ for $C^S_{RR} = C^V_{RR} = C^T_{RR} = 1$. The light orange band shows the Belle measurement at $1\sigma$.}
	\label{fig:afb}
\end{figure}

\begin{table}[h]
	\begin{center}
		\begin{tabular}{c|c|c c c| c c c}
			\hline\hline
			& $m_N$~(GeV) & $C^V_{RR}$ & $C^S_{RR}$ & $C^T_{RR}$& $C^V_{LL}$& $C^S_{LL}$& $C^T_{LL}$  \\
			\hline
			BP1 & 0.4 & 0.82 & 0.1 & 0.02 & -0.4 & 0 & 0 \\
			BP2   & 1.6 & 0.15 & -0.3 & 0.06 & 0 & 0 & 0   \\
			BP3  & 0 & 0 & 0 & 0 & 0 & 0.06 & 0.02\\
			\hline\hline
		\end{tabular}
	\end{center}
	\caption{The parameters for three benchmark points. The WCs not listed are zero.}
	\label{tab:bmp}
\end{table}

\begin{figure*}[t]
	\centering
	\includegraphics[width=0.3\textwidth]{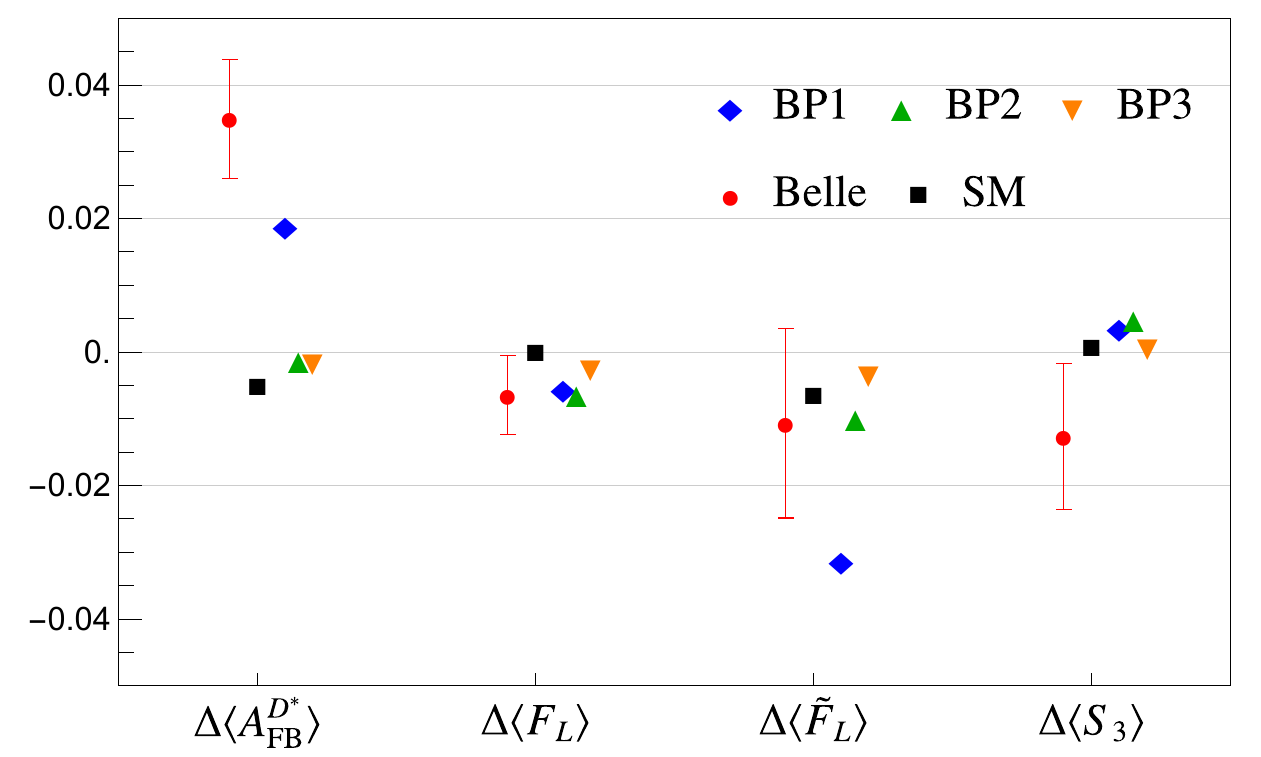}\,
	\includegraphics[width=0.3\textwidth]{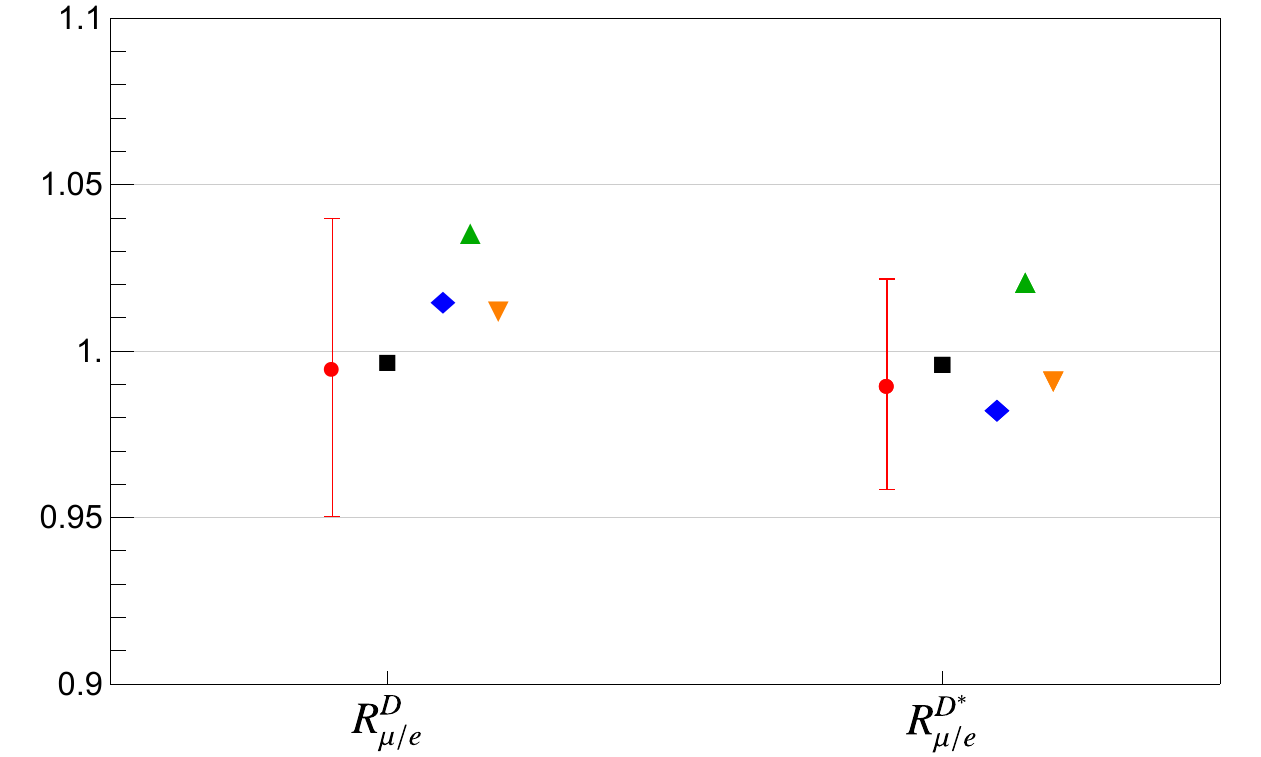}\,
	\includegraphics[width=0.3\textwidth]{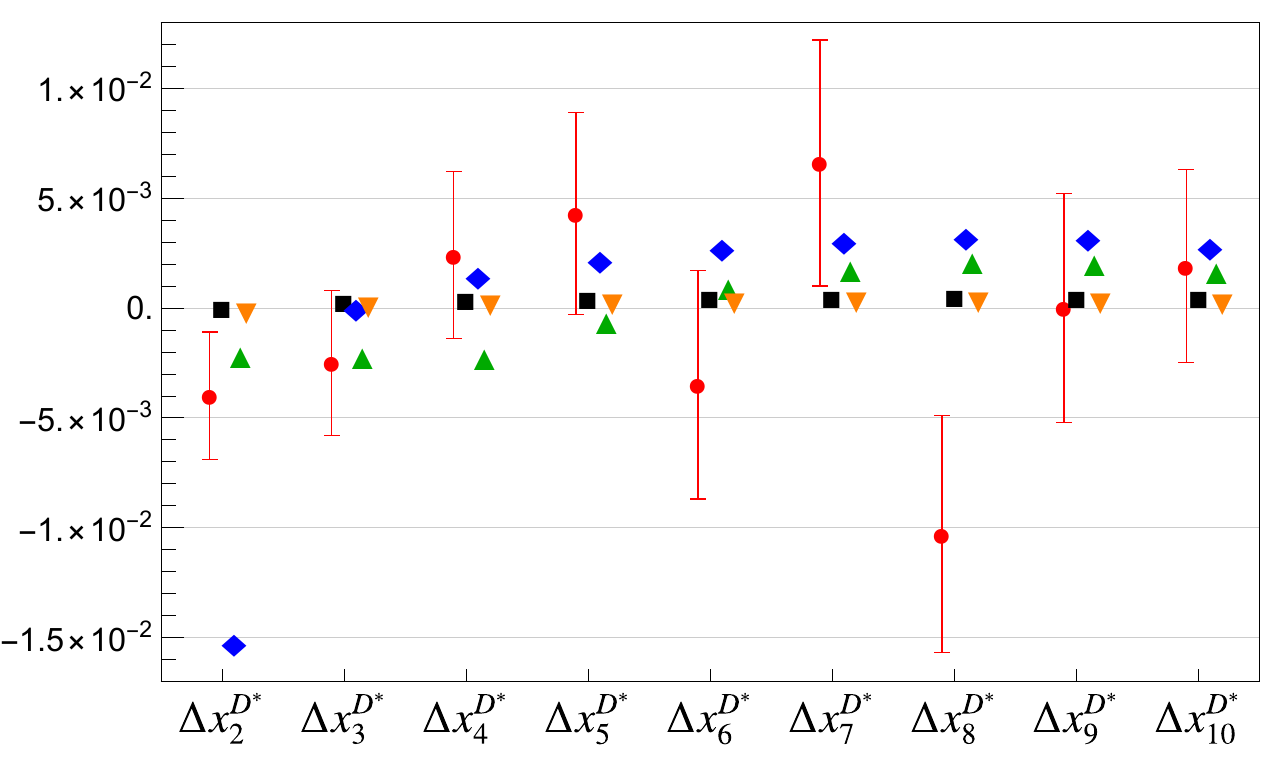}\,
	\includegraphics[width=0.3\textwidth]{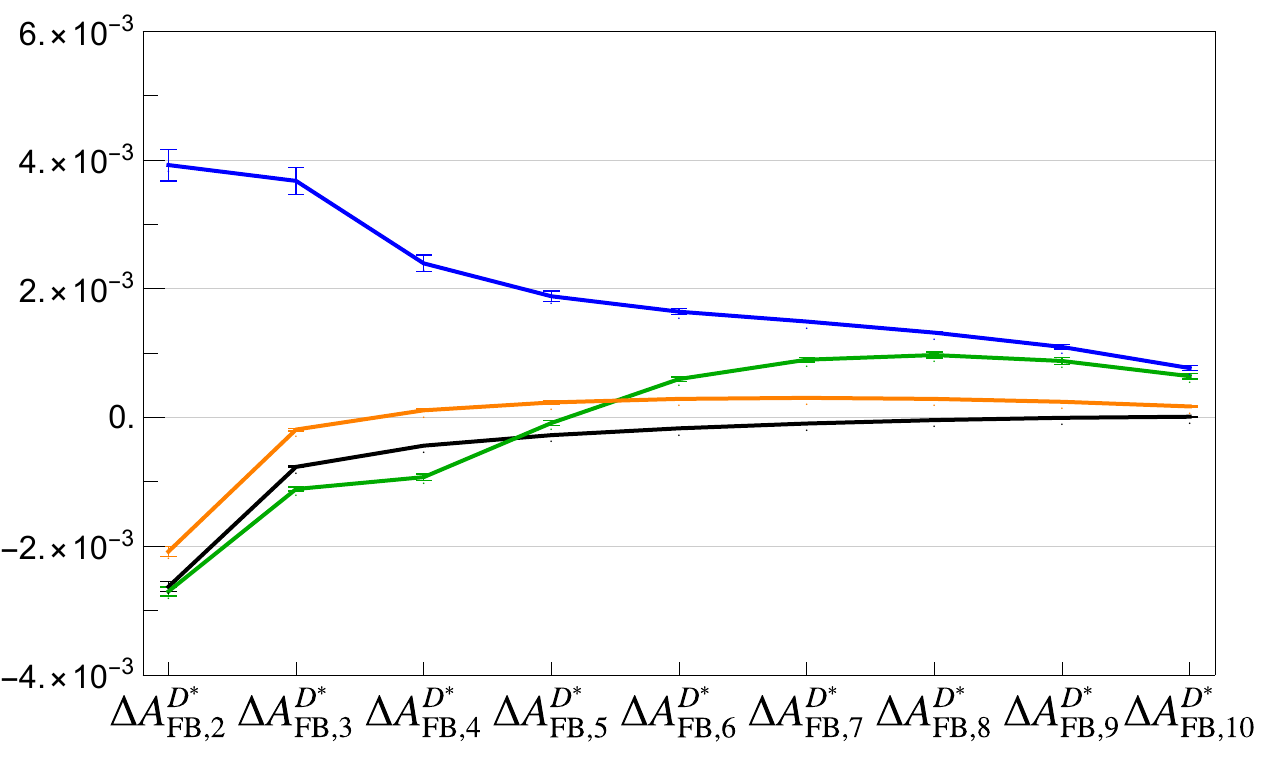}\,
	\includegraphics[width=0.3\textwidth]{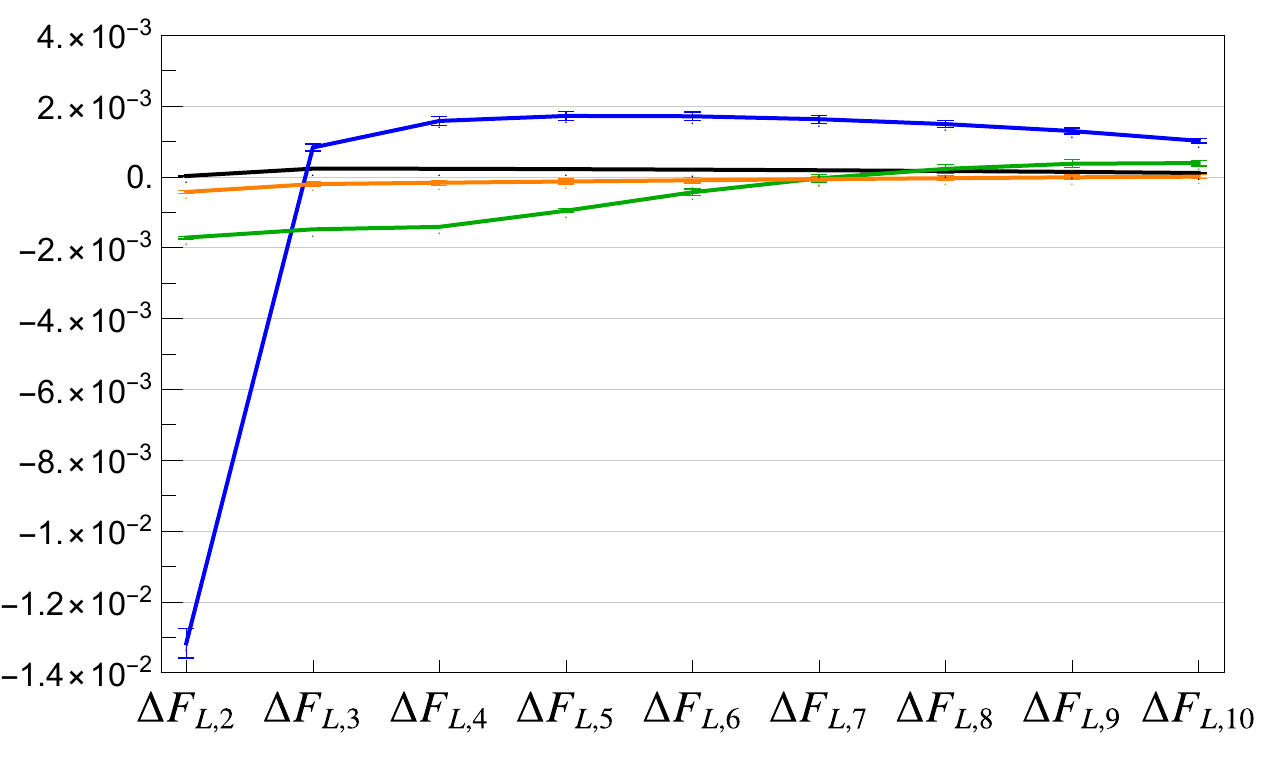}\,
	\includegraphics[width=0.3\textwidth]{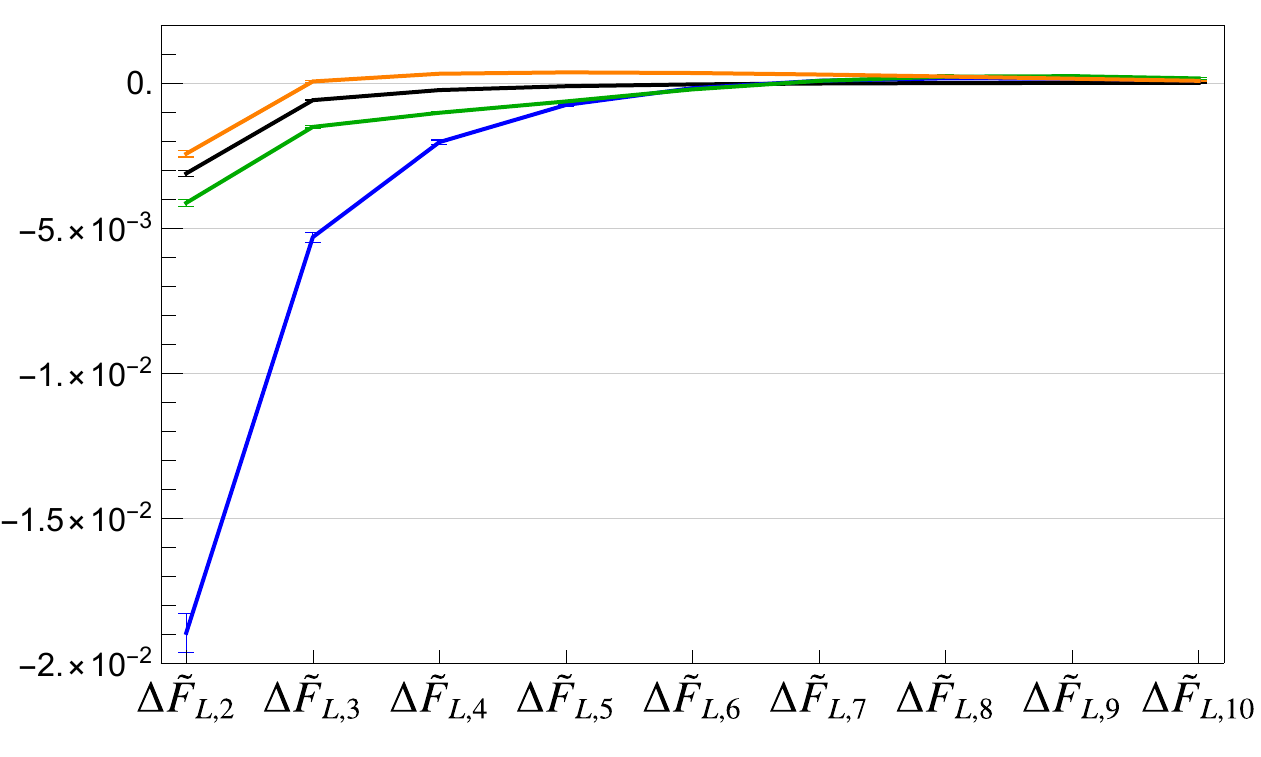}\,
	\includegraphics[width=0.3\textwidth]{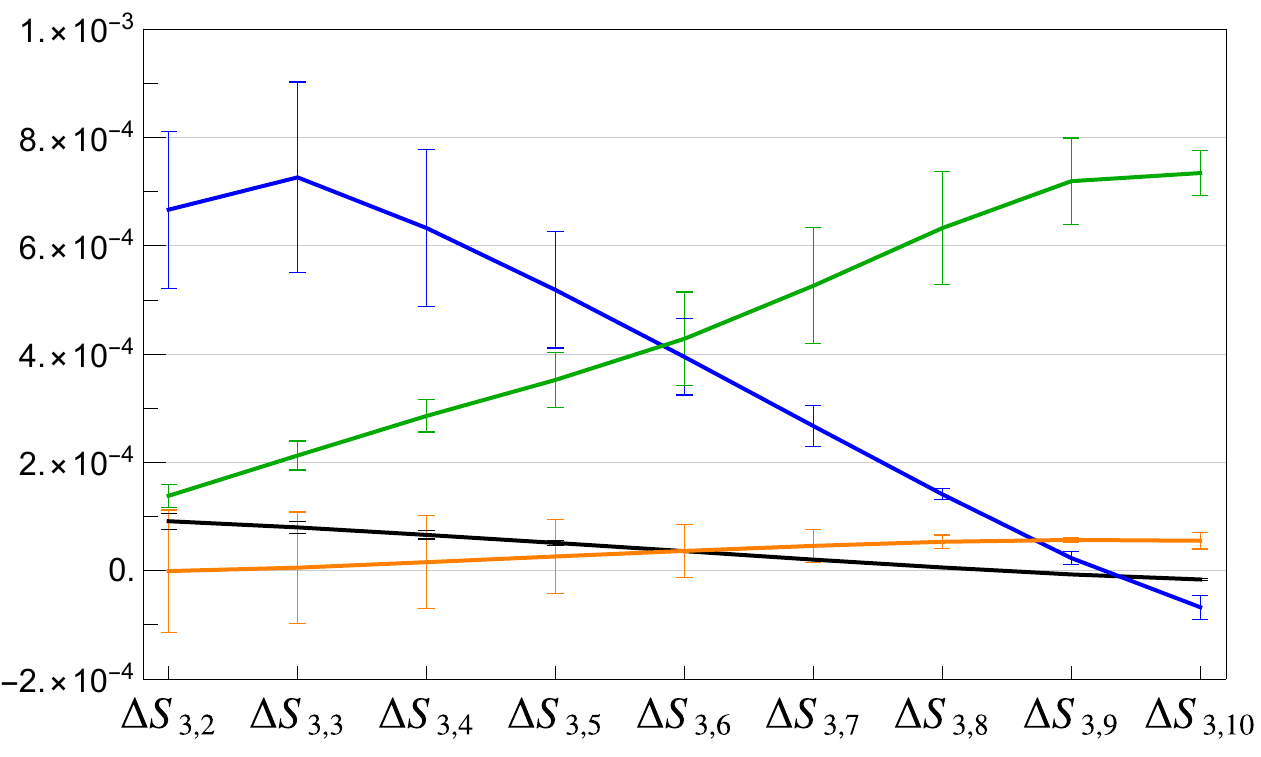}\,
	\includegraphics[width=0.3\textwidth]{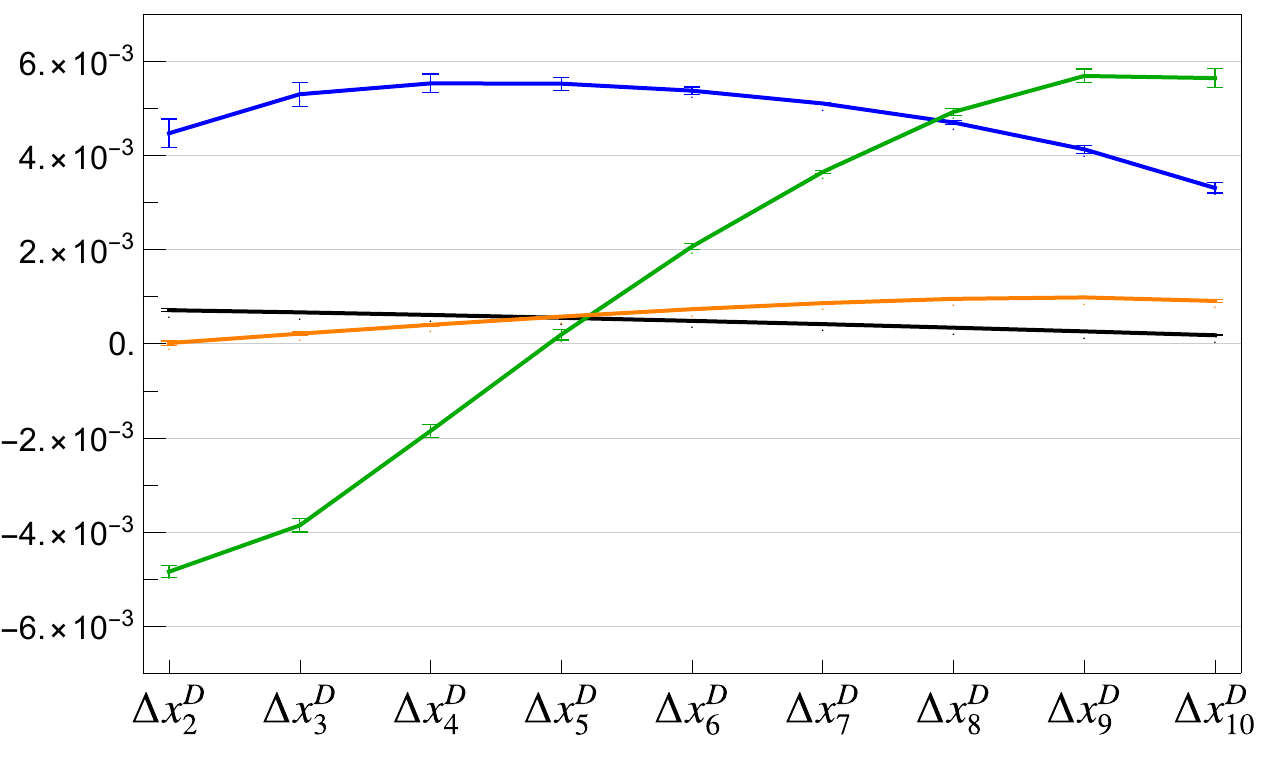}\,
	\caption{The expectations for the SM and three BPs for the observables in the upper and middle panels of Table~\ref{tab:obs_meas}. The Belle measurements are shown in the upper panels. The error bars in the middle and lower panels are hadronic form factor uncertainties. }
	\label{fig:obs}
\end{figure*}

We find that the nonzero RH neutrino mass produces significant effects in the angular observables which may explain the $4\sigma$ tension in $\Delta \langle A^{D^*}_{\text{FB}}\rangle$. In Fig.~\ref{fig:afb}, we show $\Delta \langle A^{D^*}_{\text{FB}}\rangle$ as a function $m_N$ for $C^S_{RR} = C^V_{RR} = C^T_{RR} = 1$.  Clearly, a GeV RH neutrino with vector or tensor interactions can fit the $\Delta \langle A^{D^*}_{\text{FB}}\rangle$ measurement within 1$\sigma$. We observe that if the RH neutrino is massless, $\Delta \langle A^{D^*}_{\text{FB}}\rangle$ is below the SM prediction. However, for $m_N \simeq 1$~GeV, 
the $\Delta \langle A^{D^*}_{\text{FB}}\rangle$ anomaly can be explained if $C^V_{RR} =1$ (red curve). For $m_N \simeq 2$~GeV, the anomaly can be explained by $C^T_{RR} = 1$. However, these illustrative scenarios are excluded by other measurements in Table~\ref{tab:obs_meas}. So, to reproduce the $\Delta \langle A^{D^*}_{\text{FB}}\rangle$ anomaly and the other measurements in Table~\ref{tab:obs_meas}, we choose three benchmark points (BPs) of Table~\ref{tab:bmp}.  BP1 has both LH and RH interactions. while BP2 and BP3 only have RH and LH interactions, respectively. The predictions for the three BPs for the 15 measurements are provided in Table~\ref{tab:obs_meas} and Fig.~\ref{fig:obs}. 
Since there is no interference between LH and RH contributions, scenarios with only RH interactions (like BP2) necessarily increase $R_{D}^{\mu/e}$ and $R_{D^*}^{\mu/e}$, and it is not possible to sufficiently enhance 
$\Delta \langle A^{D^*}_{\text{FB}}\rangle$. Only LH interactions (BP3) are unable to adequately reproduce all the measurements. 

\begin{figure*}[t]
	\centering
	\includegraphics[width=0.3\textwidth]{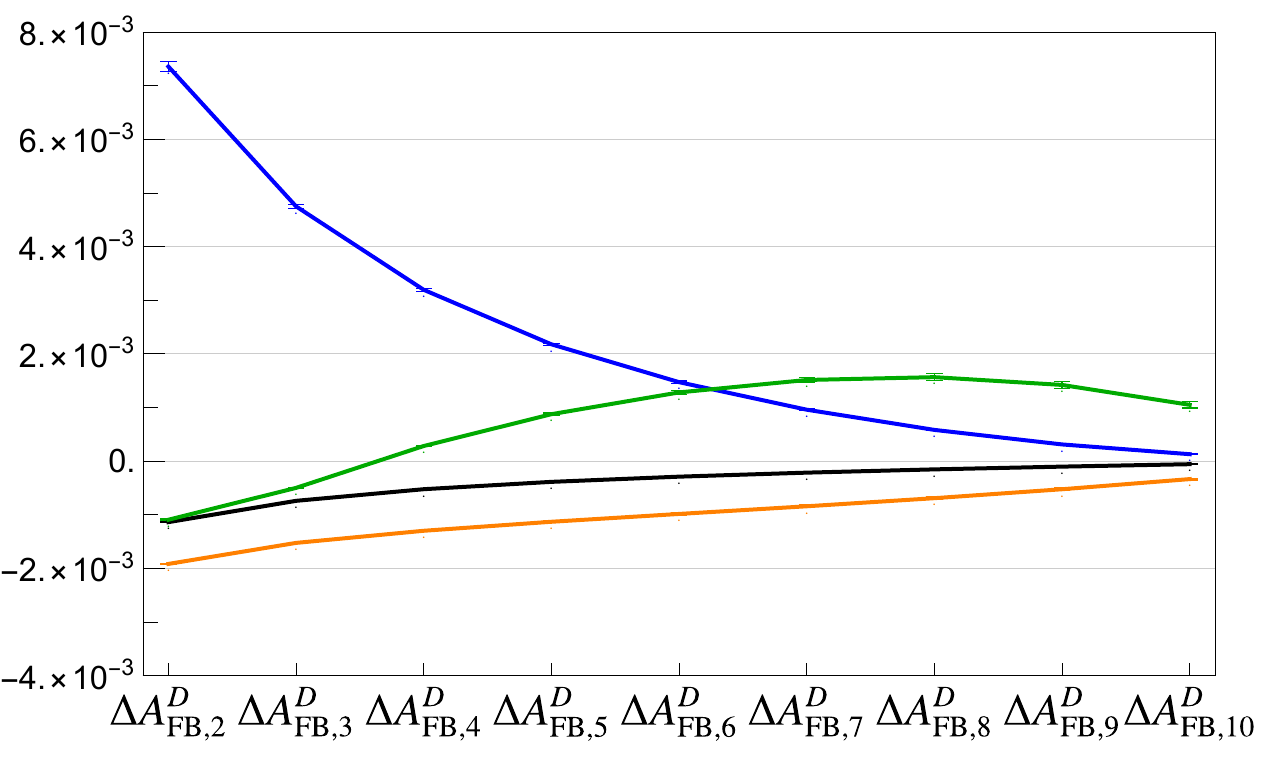}\,
	\includegraphics[width=0.3\textwidth]{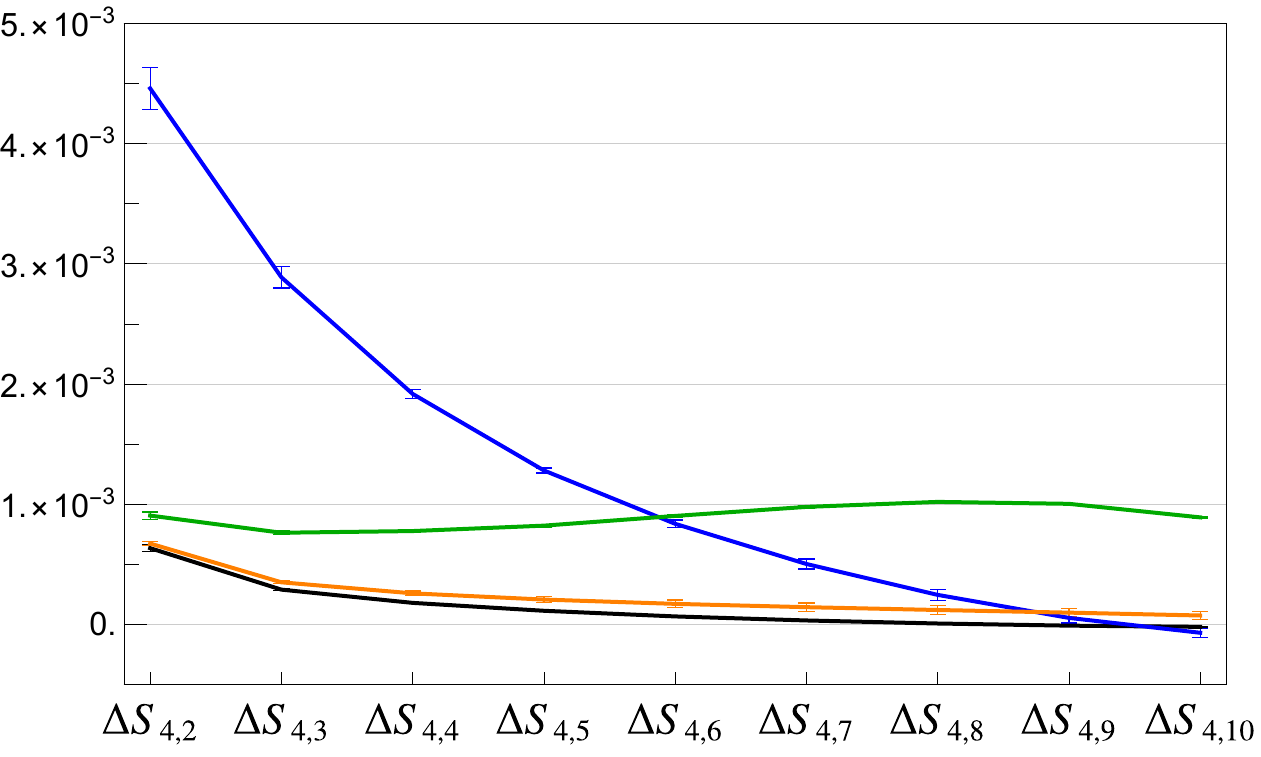}\,
	\includegraphics[width=0.3\textwidth]{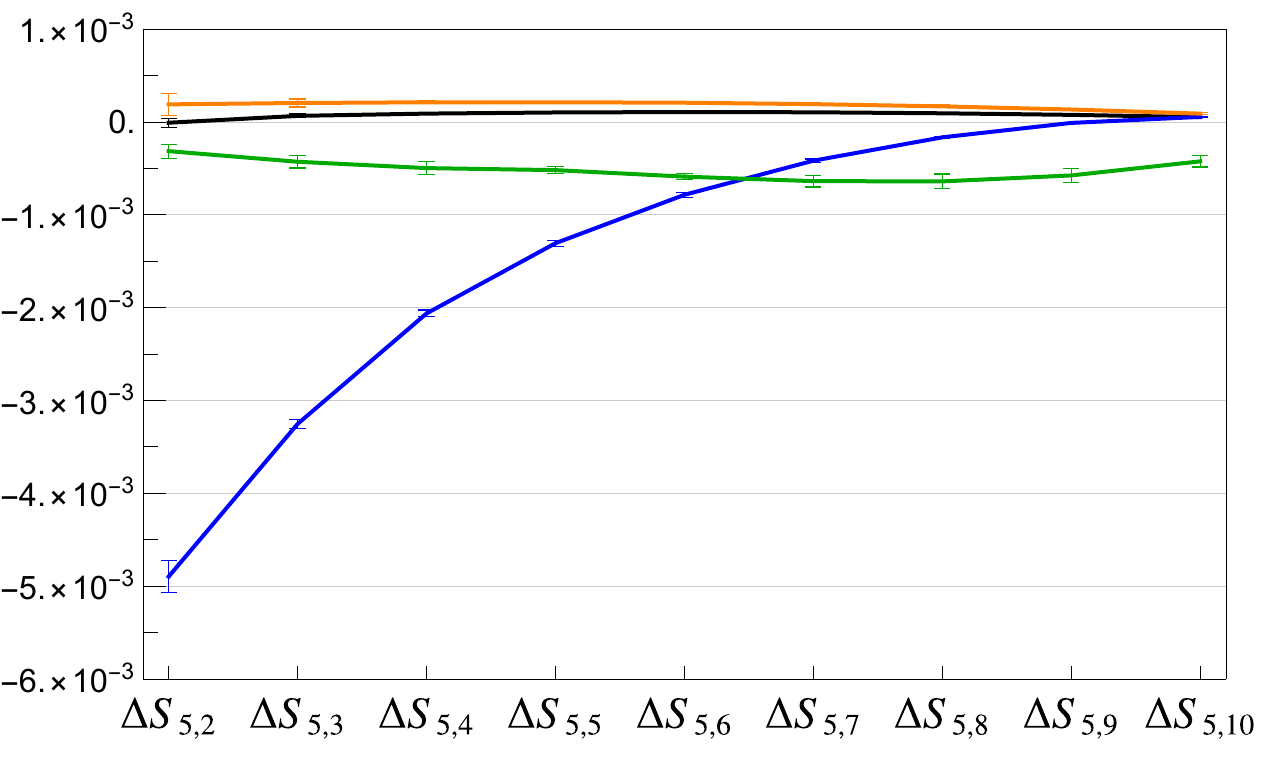}\,
	\caption{$q^2$ distributions of the three angular observables in the lower panel of Table~\ref{tab:obs_meas}.}
	\label{fig:pred2}
\end{figure*}

We now calculate $A^{D^*}_{\text{FB}}(q^2),  \tilde{F}_L(q^2),  F_L(q^2)$ and $ S_3(q^2)$ for our BP scenarios. 
We present the four binned angular observables for the three BPs in Fig.~\ref{fig:obs}. We also show the normalized $q^2$ distribution for $\bar B\ra D\ell\bar X$. 
Large deviations from the SM are evident in several $q^2$ bins.
The error bars in the middle and lower panels indicate the uncertainties due to the hadronic form factors. 
We estimate these as the range of predictions using our chosen form factors~\cite{Bordone:2019guc} and the  form factors of Refs.~\cite{Tanaka:2012nw, Iguro:2020cpg}.  We see that $\Delta S_3$ is quite sensitive to the form factor.

Other observables that have not yet been measured and can be significantly modified by NP include the forward-backward asymmetry in $\bar B\ra D \ell \bar X$,  $A^D_{\text{FB}}$.
In the SM, this is suppressed by $m_{\ell}^2$. 
The $q^2$ averaged values of $\Delta A^D_{\text{FB}}, \Delta S_4$ and  $\Delta S_5$ for the BPs are displayed in Table~\ref{tab:obs_meas}.
In Fig.~\ref{fig:pred2}, we plot the corresponding $q^2$ binned observables and find that large deviations from the SM are possible.

\section{Summary}
We find that a nonzero $m_N$ is needed to obtain a positive value of $\Delta \langle A^{D^*}_{\text{FB}}\rangle$, as suggested by Belle data. We also made predictions for several angular observables that differ substantially from SM expectations.

\begin{acknowledgments}
I thank Alakabha Datta and Danny Marfatia for collaboration on Ref.~\cite{Datta:2022czw}. This work is supported by ISF, BSF and Azrieli foundation.
\end{acknowledgments}

\bigskip 
\bibliography{Nref.bib}


\end{document}